\newcommand*\diff{\mathop{}\!\mathrm{d}}
\newcommand*\Diff[1]{\mathop{}\!\mathrm{d^#1}}
\newcommand{\vect}[1]{\boldsymbol{#1}}
\newcommand{\Ai}{{\rm Ai}}

\documentclass[%
 reprint,
showpacs,
 amsmath,amssymb,
 aps,pra
]{revtex4-1}

\usepackage{graphicx}
\usepackage{dcolumn}
\usepackage{bm}

\usepackage{nth}

\begin{document}
\title{Tunable photonic elements at the surface of an optical fiber with piezoelectric core}

\author{A. V. Dmitriev}
\email{a.dmitriev@aston.ac.uk}
\author{M. Sumetsky}
\email{m.sumetsky@aston.ac.uk}
\affiliation{Aston Institute of Photonic Technologies, Aston University, Birmingham B4 7ET, UK}

\date{\today}



\begin{abstract}
Tunable photonic elements at the surface of an optical fiber with piezoelectric core are proposed and analyzed theoretically. These elements are based on whispering gallery modes whose propagation along the fiber is fully controlled by nanoscale variation of the effective fiber radius, which can be tuned by means of a piezoelectric actuator embedded into the core. The developed theory allows one to express the introduced effective radius variation through the shape of the actuator and the voltage applied to it. In particular, the design of a miniature tunable optical delay line and a miniature tunable dispersion compensator is presented. The potential application of the suggested model to the design of a miniature optical buffer is discussed.
\end{abstract}
\maketitle

\section{Introduction}
Miniature optical delay lines, dispersion compensators and more general photonic signal processing devices are considered to be of primary importance for the future optical communications, computing, and other applications. Various implementations of microscopic optical signal processing devices have been proposed, including those based on photonic crystal structures~\cite{baba2008slow,huo2011experimental}, coupled ring resonator arrays~\cite{xia2007ultracompact} and nonlinear effects in optical fibers, such as stimulated Brillouin or Raman scattering~\cite{thevenaz2008slow}. The practical realization of miniature optical signal processing devices is however obstructed by insufficient precision and high optical losses~\cite{burmeister2008comparison,bogaerts2014design}.

Another possible approach is referred to as SNAP platform~\cite{sumetsky2011surface} and provides precise control of propagation of high-$Q$ optical whispering gallery modes (WGMs) along an optical fiber by introducing nanoscale effective fiber radius variation (ERV). Using focused CO$_2$ laser radiation to modulate the effective fiber radius by local annealing one can achieve sub-angstrom precision in the shape of fabricated photonic elements. However this technique allows one to fabricate only static surface photonic circuits and does not provide any mechanisms for the shape control and fine tuning of created devices.

SNAP circuits can be created not only by local annealing of the optical fiber, but also in several reversible ways --- these can include possible tradeoff of the precision for reconfigurability and transience. The application field of such methods is not necessarily restricted to the creation of photonic circuits from scratch (i.e. uniform optical fiber), but can include fine tuning of the pre-created "true" SNAP elements either, or provide gates for them. One of these methods exploits thermal expansion of the fiber material due to heating (to the temperatures much lower than the annealing temperature) caused by low-power focused laser radiation~\cite{dmitriev2015transient}. Another possible approach is to introduce ERV elastically, by creation of additional strain field in the vicinity of the fiber surface. Since WGMs are concentrated in a relatively narrow region near the fiber surface, the strain can be introduced by electrical control of the shape of a piezoelectric core actuator inserted into a hollow silica fiber without affecting the optical $Q$-factor.

The purpose of this paper is the theoretical analysis of this approach. The axisymmetric case of the radially polarised actuators is considered. The developed theory allows one to analytically express the introduced ERV through the shape of the actuator and the voltage applied to it. In particular, we present the design of a miniature tunable optical delay line and a miniature tunable dispersion compensator. The potential non-stationary generalisation of the suggested model that can be applied to the design of a miniature optical buffer is discussed.

\section{Theoretical formulation}
\begin{figure}[htb]
\centering
\includegraphics[width=\linewidth]{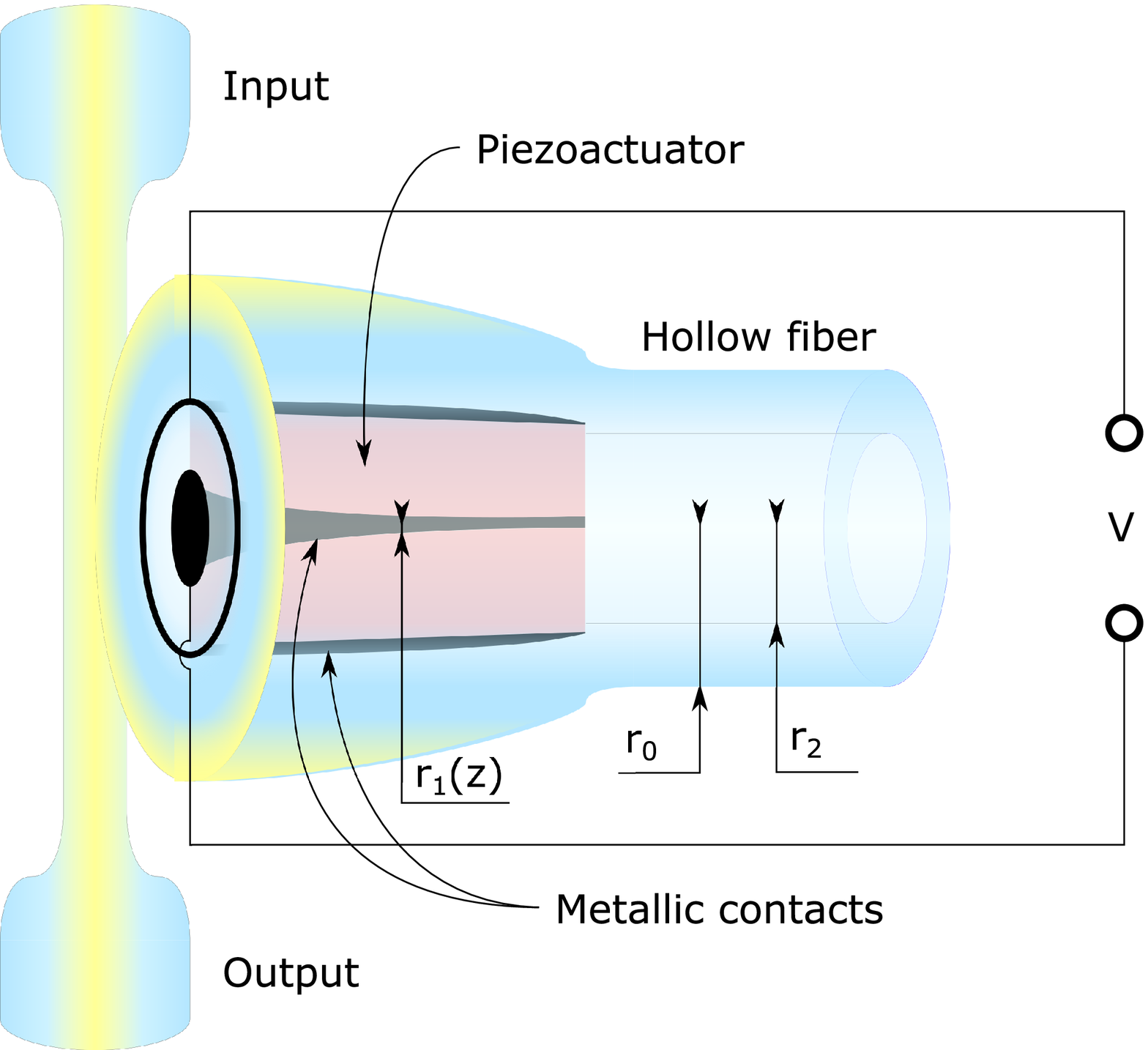}
\caption{Illustration of the tunable SNAP device considered in the paper. Voltage $V$ is applied to the radially polarized cylindrical piezoelectric transducer, which is inserted into a hollow silica fiber. The introduced strain creates nanoscale effective radius variation of the fiber along the axial coordinate, allowing one to control the propagation of whispering gallery modes in the axial direction. The device is coupled to an external photonic circuit by means of a tapered microfiber.}
\label{fig1}
\end{figure}

The considered device is illustrated in Fig.~\ref{fig1}. A piezoelectric tube-shaped actuator has the inner radius $r_1$ and outer radius $r_2$, both of which can in general vary as functions of axial coordinate $z$, metallic core and metallic coating deposited on its outer surface. The actuator is inserted into a hollow fused silica fiber with outer radius $r_0$. It is suggested that the piezoelectric material in the actuator is polarized radially, so that the axial symmetry of the structure is maintained in the process of deformation. Voltage $V$ applied to the electrodes creates strain which leads to the nanoscale deformation of the fiber shape. In addition, the strain introduces anisotropic variation of the material permittivity. Both these phenomena affect the propagation of the WGMs along the fiber. 

In this paper we consider only static deformations, i.e. the decay time of light is expected to be much smaller than characteristic time at which the fiber changes its shape. It is also assumed that the variation of the parameter values along $z$~coordinate is small and smooth enough. In this case in the vicinity of the resonant frequency the dependence $\Psi$ of the WGM electromagnetic amplitude $\vect E$ on $z$ can be factored out of the governing equations, i.e. $\vect E(r,\varphi, z) = \Psi(z) \vect \Xi(r) e^{im\theta}$. Here $\vect \Xi(r)$ describes the radial part of the electromagnetic field amplitude and $m$ is the azimuthal mode index along the coordinate $\theta$. The propagation of WGMs along $z$~axis can be described~\cite{SumetskyFiniOE2011} by a one-dimensional equation
\begin{equation}
 \frac{\Diff2 \Psi}{\diff z^2}+\beta_0^2 \left(\frac{\Delta r(z)}{r_0}+\frac{\Delta n_{eff}(z)}{n_0} - \frac{\lambda-\lambda_0-i\gamma_0}{\lambda_0} \right)\Psi=0.
\end{equation}
Here $\lambda_0$ and $\gamma_0$ are respectively the initial resonant wavelength and the loss factor of the WGM, $\beta_0=(2\pi n_0/\lambda_0)$ is the propagation constant along $z$ axis for a constant radius $r_0$, $n_0=\sqrt{\varepsilon}$ is the refractive index for non-stressed silica. To describe the dependence of WGM propagation along $z$ one can introduce dimensionless effective radius variation (ERV) $\Delta\rho_{eff}(z)$:
\begin{equation}
 \Delta \rho_{eff}(z)= \frac{\Delta r(z)}{r_0} + \frac{\Delta n_{eff}(z)}{n_0}, \label{ERV}
\end{equation}
which incorporates the geometrical variation of the outer radius $\Delta r(z)$ and the effective refractive index variation $\Delta n_{eff}(z)$.

The first term in (\ref{ERV}) can be expressed as radial mechanical displacement component at the fiber surface, while the second one describes the influence of the anisotropic perturbation of the permittivity tensor at each point on the propagation of WGMs at this point and can therefore be obtained from the first order perturbation theory~\cite{kottke2008}:
\begin{equation}
 \Delta n_{eff}(z)=-n_0\frac{\iint_{\sigma}\vect{E_0}^*\cdot\Delta\varepsilon(z) \vect{E_0} \Diff2\vect{x}}{2\iint_{\sigma}\left|\vect E_0\right|^2 \Diff2\vect{x}},\label{wavelRV}
\end{equation}
where $\sigma$ is the cross section of the hollow silica fiber and $\vect E_0$ is the WGM shape for the unperturbed structure. The first-order variation of the permittivity tensor $\Delta\varepsilon$ is coupled to the strain tensor $S$ via the photoelastic tensor $p$:
\begin{equation}
 \Delta\varepsilon=-\varepsilon^2(\Delta\varepsilon)^{-1}=-\varepsilon^2 p \vect{S}\label{perm}
\end{equation}
Here $p$ is in general a \nth{4} rank tensor, but due to the symmetry of the strain and permittivity tensors can be reduced to a \nth{2} rank tensor represented by a $6\times 6$ matrix using Voigt notation.

The electromechanical problem of piezoelectricity in the considered case is axisymmetric. Since the characteristic distance of significant change of the shape is much greater than the fiber's diameter, a plane strain approximation along $z$ axis can be used. This means that the only non-zero component of the dimensionless (in units of $r_0$) displacement vector is $u_{\rho}(\rho)$, where $\rho=r/r_0$, the only non-zero components of the strain tensor are $S_{rr}=\partial u_\rho/\partial\rho$ and $S_{\theta\theta} = u_\rho/\rho$. The equation of motion can be written down as~\cite{dianov1970analysis}
\begin{equation}
 \frac{\partial T_{rr}}{\partial\rho}+\frac1\rho \left(T_{rr}-T_{\theta\theta}\right)=0.\label{EqMoStress}
\end{equation}
In the piezoelectric region ($\rho_1<\rho<\rho_2$, $\rho_{1,2}=r_{1,2}/r_0$) the stress tensor $\vect T$ is coupled to the electric displacement $\vect D$ by constitutive equations~\cite{craig2011fundamentals}:
\begin{align}
 \vect T &=c^E \vect S + e^t \vect \nabla\varphi, \label{stress}\\
 \vect D &= e \vect T- \varepsilon^S \vect\nabla\varphi.\label{Delec}
\end{align}
Here $\varphi$ is the electric potential, $e$ is stress-charge coupling matrix (which is non-zero inside the piezoelectric only) with superscript $t$ denoting matrix transposition, while $c^{\vect E}$ and $\varepsilon^{\vect S}$ are the elasticity and permittivity matrices at constant electric and constant strain fields, respectively.

As long as the radial electric displacement in the piezoelectric is inverse proportional to the radial coordinate, $D_r=A/\rho$, one can write explicitly
\begin{align}
 T_{rr}=c_{33}^E \frac{\partial u_\rho}{\partial\rho}+c_{13}^E\frac{u_{\rho}}{\rho}+e_{33}\frac1{r_0}\frac{\partial\varphi}{\partial\rho}, \label{stressRR}\\
 T_{\theta\theta}=c_{13}^E \frac{\partial u_\rho}{\partial\rho}+c_{11}^E\frac{u_{\rho}}{\rho}+e_{31}\frac1{r_0}\frac{\partial\varphi}{\partial\rho},\label{stressTT}
\end{align}	
and
\begin{equation}
 \varphi(\rho)= \frac1{\varepsilon_{33}^S}\left[e_{33}u_\rho+ e_{31} \int\limits_{\rho_1}^\rho\frac{u_\rho(\rho')}{\rho'}\diff \rho' - A \ln\frac{\rho}{\rho_1}\right].\label{potential}
\end{equation}

The substitution of eqs. (\ref{stressRR}-\ref{potential}) into (\ref{EqMoStress}) gives
\begin{equation}
 \frac{\partial^2 u_\rho}{\partial\rho^2}+\frac1{\rho}\frac{\partial u}{\partial \rho}-\nu^2 \frac{u_\rho}{\rho^2}=-\frac{\overline{A}}{\rho^2},
\end{equation}
where 
\begin{equation*}
 \nu^2 = \frac{c_{11}^E+e_{31}^2/\varepsilon_{33}^S}{c_{33}^E+e_{33}^2/\varepsilon_{33}^2},~~~\overline{A}=\frac{e_{31}}{\varepsilon_{33}^S}\frac{A}{c_{33}^E+e_{33}^2/\varepsilon_{33}^S}.
\end{equation*}

Taking into account that the inner (metallic) and the outer (silica) cylindrical layers are assumed to be elastically isotropic and non-piezoelectric, i.e. in these regions $\nu=1$ and $\overline{A}=0$, and that the solution has to be finite at $\rho=0$, one can write down the solution as
\begin{equation}\label{displacement}
u_\rho(\rho)=\left\{ \begin{array}{lr}
u_\rho^I(\rho)=\alpha_1^I \rho, & 0 \le \rho<\rho_1;\\
u_\rho^{II}(\rho)=\alpha_1^{II}\rho^\nu+\alpha_2^{II}\rho^{-\nu}+\overline{A}\nu^{-2}, & \rho_1 \le\rho<\rho_2;\\
u_\rho^{III}(\rho)=\alpha_1^{III}\rho+\alpha_2^{III}\rho^{-1}, &\rho_2\le\rho\le 1.
\end{array}\right.
\end{equation}

The six constants $\alpha_{1,2}^{I,II,III}$ and $A$ are obtained from the linear algebraic system of equations defined by six boundary conditions. Four of them are given by the persistence of the radial displacement $u_r$ and radial stress tensor component $T_{rr}$ at material interfaces $\rho=\rho_1$ and $\rho=\rho_2$, one comes from the equality of the $T_{rr}$ to zero at the fiber surface $\rho_f$ and the last one comes from the electrostatics, which demands that the difference between the values of the electric potential $\varphi$ at $\rho_2$ and $\rho_1$ should be equal to the applied voltage $V$:
\begin{subequations}\begin{align}
 u_\rho^I(\rho_1)&=u_\rho^{II}(\rho_1),\label{boundcondfirst}\\
 u_\rho^{II}(\rho_2)&=u_\rho^{III}(\rho_2),\\
 T_{\rho\rho}^{I}(\rho_1)&= T_{\rho\rho}^{II}(\rho_1),\\
 T_{\rho\rho}^{II}(\rho_2) &=T_{\rho\rho}^{III}(\rho_2),\\
 T_{\rho\rho}^{III}(1)&=0,\\
 \varphi(\rho_2)-\varphi(\rho_1)&=V.\label{boundcondlast}
\end{align}\end{subequations}

The above formulations remain applicable if the shape of the piezoelectric actuator changes slowly along $z$ axis of the actuator. In this case the problem of calculation of shape of the piezoelectric actuator that is required to obtain the desired ERV $\Delta r_{eff}(z)$ amounts to expressing the ERV in terms of material and geometrical parameters, and then inverting the derived analytic expression with respect to the geometrical parameters that are supposed to change along $z$ axis. This makes the proposed method much more effective and universal as compared to the numerical simulation approach.

\section{Results and discussion} 
In the model considered in this paper it is expected that the piezoelectric actuators have the shape of cylindrical tubes made of PZT-5H ceramics with aluminium core. The outer radius $r_2$ of the actuator is assumed to be constant, while the inner one, $r_1$, varies with $z$ coordinate.

The resonant wavelength variation depends on the polarization of the electromagnetic field in the optical mode and on the radial component of the mode shape. The shape of the unperturbed modes near the fiber surface can be expressed using the Airy function approximation, i.e.
\begin{equation}
 E(r,\theta) \sim \Ai\left(\left[\frac{2 \beta_0^2}{r_0}\right]^{1/3}(r_0-r)-t_n\right)e^{im\theta},\label{Airy}
\end{equation}
where $\beta_0=2\pi n_0/\lambda_0$, $n_0=1.46$ is the refractive index of fused silica, $\lambda_0=1.55~\mu$m is the resonant wavelength and $t_n$ is the $n$-th root of the Airy function ($t_0\approx 2.338$, $t_1\approx 4.088$). It is found that relative effective refractive index variation $\Delta n_{eff}/n_0$ for TE$_{m0}$ and TM$_{m0}$ modes are of same order of magnitude as the relative change of the geometrical fiber radius, $\Delta r/r_0$, and therefore neither of these effects is negligible when calculating the the total value of ERV. The further analysis presented in this paper is constrained to the case of TM$_{m0}$ modes only, i.e. it is assumed that the only non-zero component of the electric field is the radial one $E_r$, and that it can be described by (\ref{Airy}) with $t_n=t_0$.

\begin{figure}[hbt]
\centering
\includegraphics[width=\linewidth]{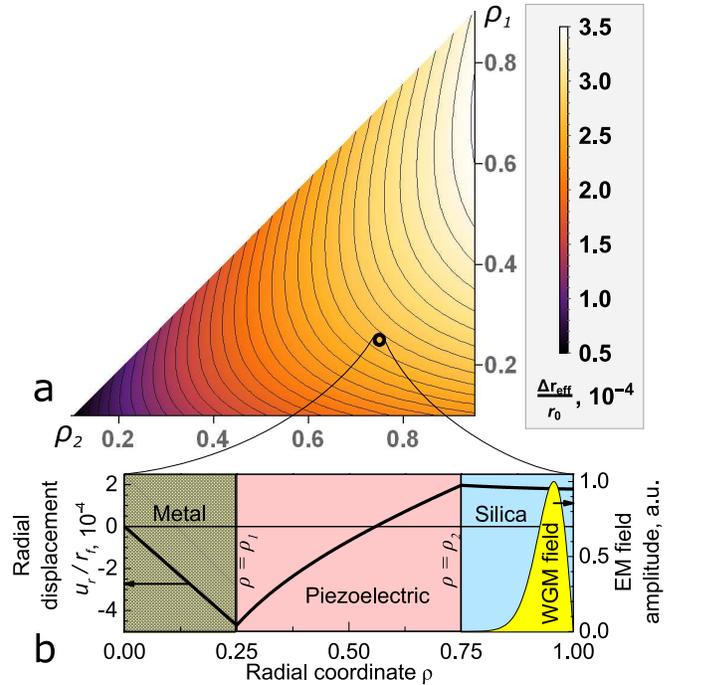}
\caption{(a) The distribution of the effective radius variation (ERV) for the fundamental TM mode as a function of dimensionless geometrical parameters $\rho_1=r_1/r_0$ and $\rho_2=r_2/r_0$; (b) The typical distribution of the radial displacement $u_\rho$ along $r$ axis for $\rho_1=0.25$, $\rho_2=0.75$, $r_0=20$~$\mu$m, $V= 100$~V (thick solid line, left scale); the radial distribution of the electric field in WGM with $\lambda_0=1.55~\mu$m, filled region, right scale. }
\label{fig2}
\end{figure}

The calculations are performed in two steps. At first, the system of equations (\ref{boundcondfirst}-\ref{boundcondlast}) is solved analytically to obtain the coefficients $\alpha_{1,2}^{I,II,III}$ for the radial displacement (\ref{displacement}). This allows one to express the ERV (\ref{ERV}) in terms of material and spatial parameters after calculating effective refractive index variation through eq.~(\ref{wavelRV}). The permittivity tensor is calculated from strain by eq.~(\ref{perm}). Since silica is an initially isotropic material, its photoelastic tensor $p$ has only two independent components, $p_{11}=0.121$ and $p_{12}=0.27$~\cite{dixon1967photoelastic}. Second, the shape of the piezoelectric actuator that is required to obtain the desired ERV distribution along $z$, $\Delta r_{eff}(z)$, is determined. To do this one needs to fix the outer actuator radius $r_2$ and express the inner radius $r_1$ as a function of $z$, i.e. $r_1=r_1(\Delta r_{eff}(z))$.

The distribution of the ERV as a function of dimensionless geometrical parameters $\rho_1=r_1/r_0$ and $\rho_2=r_2/r_0$ is given in Fig.~\ref{fig2}a. The voltage to fiber radius ratio is $V/r_0=5\times 10^{6}$~V/m. The typical values of the introduced ERV are of the order of $10^{-4}$ and for a given $\rho_2$ there is an optimal value of $\rho_1$ for which the resulting ERV is maximal.

The distribution of the radial displacement component along dimensionless radial coordinate $\rho$ for $\rho_1=0.25$, $\rho_2=0.75$ is shown in Fig.~\ref{fig2}b. The same figure shows the corresponding distribution of $E_z$ component of the WGM field given by (\ref{Airy}) for $r_0=20$~$\mu$m. Notice that in order to avoid losses the thickness of the region in the vicinity of fiber surface in which the WGM power is concentrated has to be smaller than the silica layer thickness. Fig.~\ref{fig2}b testifies that the resulting values of displacement on the fiber surface significantly depend on the the properties of the core material.

\begin{figure}[hbt]
\centering
\includegraphics[width=\linewidth]{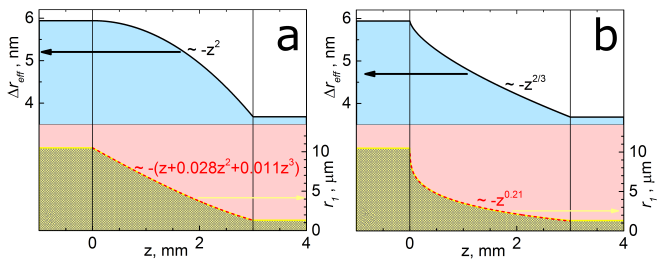}
\caption{The cross-section of a dispersionless delay line (a) and a dispersion compensator (b). Solid lines, left scales: ERV introduced into the structure. Solid lines, right scales: inner piezoelectric actuator radius $r_1$. Dashed lines, right scales: fit results for $r_1(z)$. The fiber radius is 20~$\mu$m, the outer piezoelectric actuator radius is 15~$\mu$m, applied voltage is 100~V.}
\label{fig3}
\end{figure}

The calculated distribution of the ERV is utilized to determine the piezoelectric actuator shapes for two important cases --- a dispersionless delay line and a dispersion compensator. The parabolic dependence of the ERV on $z$, leading to the dispersionless pulse propagation~\cite{sumetskyPRL}, and the corresponding piezoelectric actuator shape are shown in Fig.~\ref{fig3}a. The structure shown in Fig.~\ref{fig3}b has the ERV proportional to $z^{2/3}$, which corresponds to the constant dispersion, i.e. the linear group delay and, therefore, to the dispersion compensation behavior for incoming optical pulses~\cite{sumetsky2014management}.

An interesting opportunity is to take advantage of rapid switching abilities of piezoelectric actuators in order to construct fast switching devices such as SNAP-based harmonic optical buffers described in~\cite{sumetsky2015buffer}. A non-stationary generalization of the above analysis is required to figure out the potential characteristics of such a device. Assuming that the response rate of such buffer is primarily determined by the time required for the acoustic pulses to travel through silica from the piezoelectric actuator to the outer fiber surface, one can roughly estimate that this rate can take values of a few GHz. It is suggested that the considered model of the piezoelectric actuator with constant outer radius $r_2$ ensures that the acoustic vibrations reach the fiber surface with the same phase.

It is assumed above that the ERV in the absence of voltage applied to the electrodes is zero, i.e. the fiber is initially uniform. However, the nanoscale initial ERV won't change the above formulations. Therefore, the technique presented here can be used to add tunability to the SNAP devices introduced by non-reversible methods.

\section{Conclusion}
In summary, a new method for creation of ultralow-loss tunable surface photonic elements in silica fibers with piezoelectric core actuators has been proposed. The axisymmetric realization of this method with radially polarized actuators has been analyzed theoretically. It has been shown that it is possible to create tunable dispersionless optical delay lines and linear dispersion compensators using this method. The shapes of the actuators that form these elements have been determined.

The obtained results show that silica fibers with piezoelectric core are very promising for the creation and control of transient SNAP photonic circuits and imply the future experimental realization and further generalization of the developed theory.  

\section*{Funding}
M. Sumetsky acknowledges the Royal Society Wolfson Research Merit Award WM130110.
\bibliography{manuscript}

\begin{thebibliography}{16}%
\makeatletter
\providecommand \@ifxundefined [1]{%
 \@ifx{#1\undefined}
}%
\providecommand \@ifnum [1]{%
 \ifnum #1\expandafter \@firstoftwo
 \else \expandafter \@secondoftwo
 \fi
}%
\providecommand \@ifx [1]{%
 \ifx #1\expandafter \@firstoftwo
 \else \expandafter \@secondoftwo
 \fi
}%
\providecommand \natexlab [1]{#1}%
\providecommand \enquote  [1]{``#1''}%
\providecommand \bibnamefont  [1]{#1}%
\providecommand \bibfnamefont [1]{#1}%
\providecommand \citenamefont [1]{#1}%
\providecommand \href@noop [0]{\@secondoftwo}%
\providecommand \href [0]{\begingroup \@sanitize@url \@href}%
\providecommand \@href[1]{\@@startlink{#1}\@@href}%
\providecommand \@@href[1]{\endgroup#1\@@endlink}%
\providecommand \@sanitize@url [0]{\catcode `\\12\catcode `\$12\catcode
  `\&12\catcode `\#12\catcode `\^12\catcode `\_12\catcode `\%12\relax}%
\providecommand \@@startlink[1]{}%
\providecommand \@@endlink[0]{}%
\providecommand \url  [0]{\begingroup\@sanitize@url \@url }%
\providecommand \@url [1]{\endgroup\@href {#1}{\urlprefix }}%
\providecommand \urlprefix  [0]{URL }%
\providecommand \Eprint [0]{\href }%
\providecommand \doibase [0]{http://dx.doi.org/}%
\providecommand \selectlanguage [0]{\@gobble}%
\providecommand \bibinfo  [0]{\@secondoftwo}%
\providecommand \bibfield  [0]{\@secondoftwo}%
\providecommand \translation [1]{[#1]}%
\providecommand \BibitemOpen [0]{}%
\providecommand \bibitemStop [0]{}%
\providecommand \bibitemNoStop [0]{.\EOS\space}%
\providecommand \EOS [0]{\spacefactor3000\relax}%
\providecommand \BibitemShut  [1]{\csname bibitem#1\endcsname}%
\let\auto@bib@innerbib\@empty
\bibitem [{\citenamefont {Baba}(2008)}]{baba2008slow}%
  \BibitemOpen
  \bibfield  {author} {\bibinfo {author} {\bibfnamefont {T.}~\bibnamefont
  {Baba}},\ }\href@noop {} {\bibfield  {journal} {\bibinfo  {journal} {Nature
  photonics}\ }\textbf {\bibinfo {volume} {2}},\ \bibinfo {pages} {465}
  (\bibinfo {year} {2008})}\BibitemShut {NoStop}%
\bibitem [{\citenamefont {Huo}\ \emph {et~al.}(2011)\citenamefont {Huo},
  \citenamefont {Sandhu}, \citenamefont {Pan}, \citenamefont {Stuhrmann},
  \citenamefont {Povinelli}, \citenamefont {Kahn}, \citenamefont {Harris},
  \citenamefont {Fejer},\ and\ \citenamefont {Fan}}]{huo2011experimental}%
  \BibitemOpen
  \bibfield  {author} {\bibinfo {author} {\bibfnamefont {Y.}~\bibnamefont
  {Huo}}, \bibinfo {author} {\bibfnamefont {S.}~\bibnamefont {Sandhu}},
  \bibinfo {author} {\bibfnamefont {J.}~\bibnamefont {Pan}}, \bibinfo {author}
  {\bibfnamefont {N.}~\bibnamefont {Stuhrmann}}, \bibinfo {author}
  {\bibfnamefont {M.~L.}\ \bibnamefont {Povinelli}}, \bibinfo {author}
  {\bibfnamefont {J.~M.}\ \bibnamefont {Kahn}}, \bibinfo {author}
  {\bibfnamefont {J.~S.}\ \bibnamefont {Harris}}, \bibinfo {author}
  {\bibfnamefont {M.~M.}\ \bibnamefont {Fejer}}, \ and\ \bibinfo {author}
  {\bibfnamefont {S.}~\bibnamefont {Fan}},\ }\href@noop {} {\bibfield
  {journal} {\bibinfo  {journal} {Optics letters}\ }\textbf {\bibinfo {volume}
  {36}},\ \bibinfo {pages} {1482} (\bibinfo {year} {2011})}\BibitemShut
  {NoStop}%
\bibitem [{\citenamefont {Xia}\ \emph {et~al.}(2007)\citenamefont {Xia},
  \citenamefont {Sekaric},\ and\ \citenamefont {Vlasov}}]{xia2007ultracompact}%
  \BibitemOpen
  \bibfield  {author} {\bibinfo {author} {\bibfnamefont {F.}~\bibnamefont
  {Xia}}, \bibinfo {author} {\bibfnamefont {L.}~\bibnamefont {Sekaric}}, \ and\
  \bibinfo {author} {\bibfnamefont {Y.}~\bibnamefont {Vlasov}},\ }\href@noop {}
  {\bibfield  {journal} {\bibinfo  {journal} {Nature photonics}\ }\textbf
  {\bibinfo {volume} {1}},\ \bibinfo {pages} {65} (\bibinfo {year}
  {2007})}\BibitemShut {NoStop}%
\bibitem [{\citenamefont {Th{\'e}venaz}(2008)}]{thevenaz2008slow}%
  \BibitemOpen
  \bibfield  {author} {\bibinfo {author} {\bibfnamefont {L.}~\bibnamefont
  {Th{\'e}venaz}},\ }\href@noop {} {\bibfield  {journal} {\bibinfo  {journal}
  {Nature Photonics}\ }\textbf {\bibinfo {volume} {2}},\ \bibinfo {pages} {474}
  (\bibinfo {year} {2008})}\BibitemShut {NoStop}%
\bibitem [{\citenamefont {Burmeister}\ \emph {et~al.}(2008)\citenamefont
  {Burmeister}, \citenamefont {Blumenthal},\ and\ \citenamefont
  {Bowers}}]{burmeister2008comparison}%
  \BibitemOpen
  \bibfield  {author} {\bibinfo {author} {\bibfnamefont {E.~F.}\ \bibnamefont
  {Burmeister}}, \bibinfo {author} {\bibfnamefont {D.~J.}\ \bibnamefont
  {Blumenthal}}, \ and\ \bibinfo {author} {\bibfnamefont {J.~E.}\ \bibnamefont
  {Bowers}},\ }\href@noop {} {\bibfield  {journal} {\bibinfo  {journal}
  {Optical Switching and Networking}\ }\textbf {\bibinfo {volume} {5}},\
  \bibinfo {pages} {10} (\bibinfo {year} {2008})}\BibitemShut {NoStop}%
\bibitem [{\citenamefont {Bogaerts}\ \emph {et~al.}(2014)\citenamefont
  {Bogaerts}, \citenamefont {Fiers},\ and\ \citenamefont
  {Dumon}}]{bogaerts2014design}%
  \BibitemOpen
  \bibfield  {author} {\bibinfo {author} {\bibfnamefont {W.}~\bibnamefont
  {Bogaerts}}, \bibinfo {author} {\bibfnamefont {M.}~\bibnamefont {Fiers}}, \
  and\ \bibinfo {author} {\bibfnamefont {P.}~\bibnamefont {Dumon}},\
  }\href@noop {} {\bibfield  {journal} {\bibinfo  {journal} {Selected Topics in
  Quantum Electronics, IEEE Journal of}\ }\textbf {\bibinfo {volume} {20}},\
  \bibinfo {pages} {1} (\bibinfo {year} {2014})}\BibitemShut {NoStop}%
\bibitem [{\citenamefont {Sumetsky}\ \emph {et~al.}(2011)\citenamefont
  {Sumetsky}, \citenamefont {DiGiovanni}, \citenamefont {Dulashko},
  \citenamefont {Fini}, \citenamefont {Liu}, \citenamefont {Monberg},\ and\
  \citenamefont {Taunay}}]{sumetsky2011surface}%
  \BibitemOpen
  \bibfield  {author} {\bibinfo {author} {\bibfnamefont {M.}~\bibnamefont
  {Sumetsky}}, \bibinfo {author} {\bibfnamefont {D.}~\bibnamefont
  {DiGiovanni}}, \bibinfo {author} {\bibfnamefont {Y.}~\bibnamefont
  {Dulashko}}, \bibinfo {author} {\bibfnamefont {J.}~\bibnamefont {Fini}},
  \bibinfo {author} {\bibfnamefont {X.}~\bibnamefont {Liu}}, \bibinfo {author}
  {\bibfnamefont {E.}~\bibnamefont {Monberg}}, \ and\ \bibinfo {author}
  {\bibfnamefont {T.}~\bibnamefont {Taunay}},\ }\href@noop {} {\bibfield
  {journal} {\bibinfo  {journal} {Optics Letters}\ }\textbf {\bibinfo {volume}
  {36}},\ \bibinfo {pages} {4824} (\bibinfo {year} {2011})}\BibitemShut
  {NoStop}%
\bibitem [{\citenamefont {Dmitriev}\ \emph {et~al.}(2015)\citenamefont
  {Dmitriev}, \citenamefont {Toropov},\ and\ \citenamefont
  {Sumetsky}}]{dmitriev2015transient}%
  \BibitemOpen
  \bibfield  {author} {\bibinfo {author} {\bibfnamefont {A.}~\bibnamefont
  {Dmitriev}}, \bibinfo {author} {\bibfnamefont {N.}~\bibnamefont {Toropov}}, \
  and\ \bibinfo {author} {\bibfnamefont {M.}~\bibnamefont {Sumetsky}},\ }in\
  \href {\doibase 10.1109/IPCon.2015.7323759} {\emph {\bibinfo {booktitle}
  {Photonics Conference (IPC), 2015}}}\ (\bibinfo {year} {2015})\ pp.\ \bibinfo
  {pages} {1--2}\BibitemShut {NoStop}%
\bibitem [{\citenamefont {Sumetsky}\ and\ \citenamefont
  {Fini}(2011)}]{SumetskyFiniOE2011}%
  \BibitemOpen
  \bibfield  {author} {\bibinfo {author} {\bibfnamefont {M.}~\bibnamefont
  {Sumetsky}}\ and\ \bibinfo {author} {\bibfnamefont {J.}~\bibnamefont
  {Fini}},\ }\href@noop {} {\bibfield  {journal} {\bibinfo  {journal} {Optics
  express}\ }\textbf {\bibinfo {volume} {19}},\ \bibinfo {pages} {26470}
  (\bibinfo {year} {2011})}\BibitemShut {NoStop}%
\bibitem [{\citenamefont {Kottke}\ \emph {et~al.}(2008)\citenamefont {Kottke},
  \citenamefont {Farjadpour},\ and\ \citenamefont {Johnson}}]{kottke2008}%
  \BibitemOpen
  \bibfield  {author} {\bibinfo {author} {\bibfnamefont {C.}~\bibnamefont
  {Kottke}}, \bibinfo {author} {\bibfnamefont {A.}~\bibnamefont {Farjadpour}},
  \ and\ \bibinfo {author} {\bibfnamefont {S.~G.}\ \bibnamefont {Johnson}},\
  }\href@noop {} {\bibfield  {journal} {\bibinfo  {journal} {Physical Review
  E}\ }\textbf {\bibinfo {volume} {77}},\ \bibinfo {pages} {036611} (\bibinfo
  {year} {2008})}\BibitemShut {NoStop}%
\bibitem [{\citenamefont {Dianov}\ and\ \citenamefont
  {Kuz'menko}(1970)}]{dianov1970analysis}%
  \BibitemOpen
  \bibfield  {author} {\bibinfo {author} {\bibfnamefont {D.}~\bibnamefont
  {Dianov}}\ and\ \bibinfo {author} {\bibfnamefont {A.}~\bibnamefont
  {Kuz'menko}},\ }\href@noop {} {\bibfield  {journal} {\bibinfo  {journal}
  {Soviet Physics Acoustics}\ }\textbf {\bibinfo {volume} {16}},\ \bibinfo
  {pages} {34} (\bibinfo {year} {1970})}\BibitemShut {NoStop}%
\bibitem [{\citenamefont {Craig}\ and\ \citenamefont
  {Kurdila}(2011)}]{craig2011fundamentals}%
  \BibitemOpen
  \bibfield  {author} {\bibinfo {author} {\bibfnamefont {R.}~\bibnamefont
  {Craig}}\ and\ \bibinfo {author} {\bibfnamefont {A.}~\bibnamefont
  {Kurdila}},\ }\href {https://books.google.co.uk/books?id=RM6MYkiF-mAC} {\emph
  {\bibinfo {title} {Fundamentals of Structural Dynamics}}}\ (\bibinfo
  {publisher} {Wiley},\ \bibinfo {year} {2011})\BibitemShut {NoStop}%
\bibitem [{\citenamefont {Dixon}(1967)}]{dixon1967photoelastic}%
  \BibitemOpen
  \bibfield  {author} {\bibinfo {author} {\bibfnamefont {R.}~\bibnamefont
  {Dixon}},\ }\href@noop {} {\bibfield  {journal} {\bibinfo  {journal} {Journal
  of Applied Physics}\ }\textbf {\bibinfo {volume} {38}},\ \bibinfo {pages}
  {5149} (\bibinfo {year} {1967})}\BibitemShut {NoStop}%
\bibitem [{\citenamefont {Sumetsky}(2013)}]{sumetskyPRL}%
  \BibitemOpen
  \bibfield  {author} {\bibinfo {author} {\bibfnamefont {M.}~\bibnamefont
  {Sumetsky}},\ }\href {\doibase 10.1103/PhysRevLett.111.163901} {\bibfield
  {journal} {\bibinfo  {journal} {Phys. Rev. Lett.}\ }\textbf {\bibinfo
  {volume} {111}},\ \bibinfo {pages} {163901} (\bibinfo {year}
  {2013})}\BibitemShut {NoStop}%
\bibitem [{\citenamefont {Sumetsky}(2014)}]{sumetsky2014management}%
  \BibitemOpen
  \bibfield  {author} {\bibinfo {author} {\bibfnamefont {M.}~\bibnamefont
  {Sumetsky}},\ }in\ \href@noop {} {\emph {\bibinfo {booktitle} {CLEO: Science
  and Innovations}}}\ (\bibinfo {organization} {Optical Society of America},\
  \bibinfo {year} {2014})\ pp.\ \bibinfo {pages} {STu3N--6}\BibitemShut
  {NoStop}%
\bibitem [{\citenamefont {Sumetsky}(2015)}]{sumetsky2015buffer}%
  \BibitemOpen
  \bibfield  {author} {\bibinfo {author} {\bibfnamefont {M.}~\bibnamefont
  {Sumetsky}},\ }\href@noop {} {\bibfield  {journal} {\bibinfo  {journal}
  {Scientific Reports}\ }\textbf {\bibinfo {volume} {5}},\ \bibinfo {pages}
  {18569} (\bibinfo {year} {2015})}\BibitemShut {NoStop}%
\end{thebibliography}%
%
\end{document}